\documentclass[prl,twocolumn,showpacs]{revtex4}
\pdfoutput=1
\usepackage[english]{babel}
\usepackage{amssymb}
\usepackage{amsmath}
\usepackage{graphicx}
\usepackage{hyperref}

\begin{document}

\renewcommand{\figurename}{\textbf{Fig.}}	

\title{Paradox of Photons Disconnected Trajectories\\
		Being Located by Means of ``Weak Measurements''\\
		in the Nested Mach-Zehnder Interferometer}
	
\author{G. N. Nikolaev}
\email[e-mail: ]{nikolaev@iae.nsk.su}
\affiliation{Institute of Automation and Electrometry of SB RAS, Novosibirsk, 630090 Russia\\ Novosibirsk State University, Novosibirsk, 630090 Russia}
	
\begin{abstract}
	Recently, a scheme based on the method of weak measurements to register the trajectories of photons passing
	through a nested Mach-Zehnder interferometer was proposed [L. Vaidman, Phys. Rev. A \textbf{87}, 052104 (2013)]
	and then realized [A. Danan, D. Farfurnik, S. Bar-Ad, et al., Phys. Rev. Lett. \textbf{111}, 240402 (2013)]. Interpreting
	the results of the experiment, the authors concluded that ``the photons do not always follow continuous
	trajectories''. It is shown in this work that these results can be easily and clearly explained in terms of traditional
	classical electrodynamics or quantum mechanics implying the continuity of all possible paths of photons.
	Consequently, a new concept of disconnected trajectories proposed by the authors of work [Phys. Rev.
	Lett. \textbf{111}, 240402 (2013)] is unnecessary.
\end{abstract}


\maketitle

\section{\label{sec:Introduction}INTRODUCTION}

Quantum mechanics had deeply impact on the possibilities of measurements of
physical quantities. On one hand, it revealed the limits of measurement of a
number of physical quantities because of inherent ``quantum noise'' caused by
the probabilistic nature of quantum laws. On the other hand, the quantization
(discreteness) of certain physical quantities has made it possible to increase
significantly the accuracy and to avoid the manifestations of thermal noise of
measurements (M\"ossbauer effect, quantum Hall effect, superconductivity, etc.).
A new type of quantum measurements, the so-called ``weak measurements,'' was
proposed about three decades ago
\cite{Aharonov1988,Duck1989,Aharonov1990,Aharonov2005,Dressel2014,Struppa2014}.
Such measurements have a number of paradoxical properties: the measured ``weak
quantities'' of Hermitian operators can be significantly beyond the spectrum of
their values, weak quantities are generally complex, weak measurements of
noncommuting observables can be performed simultaneously, and ``weak
probabilities'' of projection operators can have negative values. Because of
such unusual properties, weak measurements are still under discussion. Despite
this discussion, weak measurements were realized soon after their prediction
\cite{Ritchie1991}. Many hundreds of works have been performed on this subject.
The unique properties initiated a great interest in weak measurements as
promising metrological tools. In particular, owing to a specific property of
enhancement of weak signals, weak measurements can be used to estimate small
changes in parameters such as the deviation of a beam of light and frequency,
phase, velocity, temperature, and time shifts. The complexity of weak quantities
whose real and imaginary parts can be measured simultaneously underlies methods
of direct measurement of quantum states, wavefunctions, and geometric phases
(see review \cite{Dressel2014} and references therein).

Using the concept of weak measurements, the authors of \cite{Danan2013}
performed a fine experiment in order to reveal the past of photons passing
through a nested Mach-Zehnder interferometer (Fig. \ref{fig:setup}). The results
were so unexpected that the authors arrived at the necessity of rejecting the
commonsense treatment of the past of a quantum particle. They concluded that the
past of photons is not described by a continuous trajectory or a set of possible
trajectories, as is commonly accepted in traditional quantum mechanics
\cite{Feynman2011ru}. The idea of such an experiment was proposed in
\cite{Vaidman2013a}, where it was predicted that weak measurements at the output
of the nested Mach-Zehnder interferometer should indicate both the presence of
photons in the inner Mach-Zehnder interferometer, which is tuned to destructive
interference, and the absence of them at its input and output. This conclusion
was based on the two-state vector formalism \cite{Aharonov1964,Aharonov1990} and
under the assumption that perturbation of the Mach-Zehnder interferometer in the
process of weak measurements can be neglected (which was also used when
interpreting the results of the experiment reported in \cite{Danan2013}).
However, the authors of \cite{Li2013} commenting on work \cite{Vaidman2013a}
qualitatively showed that weak measurements in the inner Mach-Zehnder
interferometer partially destroy the interference between waves passing through
it.

The surprising results of the experiment reported in \cite{Danan2013} were
actively discussed in
\cite{Wiesniak2014,Salih2014,Svensson2014a,Huang2014,Saldanha2014,Li2015,Bartkiewicz2015},
where their interpretation was criticized and various modifications of the
experiment and methods of data processing were proposed. The authors of
\cite{Danan2013} also reported the calculation of detected signals according to
the classical theory of electromagnetic waves but did not find any simple
interpretation of the signals within this approach. The authors of
\cite{Bartkiewicz2015} proposed to describe the evolution of the state of
photons within a modified scheme according to traditional quantum mechanics. A
numerical-analytical description of the discussed experiment was given in
\cite{Saldanha2014} on the basis of the paraxial optical approximation and
angular spectrum. The aim was to give a clear physical interpretation of
paradoxical results of the experiment reported in \cite{Danan2013}. However,
this aim was achieved incompletely. It was emphasized that interference of all
beams of light in the nested Mach-Zehnder interferometer is important for the
formation of a signal on the detector. This was explicitly declared in the
introduction in \cite{Saldanha2014} and was used to calculate the amplitude of
the angular spectrum at the output of the nested Mach-Zehnder interferometer in
the case of destructive tuning of the inner Mach-Zehnder interferometer.
However, this fundamental aspect was disregarded when deriving expressions for
signals on the photodetector, thus reducing the credibility of the results. The
considered feature is brightly indicated by an additional paradoxical result of
the discussed experiment: in the case of destructive tuning of the inner
Mach-Zehnder interferometer, all three signals (from mirrors $A$, $B$ and $C$)
disappear when the beam of light in the outer arm of the outer Mach-Zehnder
interferometer (from mirror $C$) is blocked. This result cannot be explained
within the treatment of the two-state vector formalism by the authors of
\cite{Danan2013}. As a whole, the results of the experiment reported in
\cite{Danan2013} were quite clearly interpreted in \cite{Saldanha2014}
(including numerical calculations), but with a number of uncertainties and
ambiguities. In particular, in the case of destructive interference in the inner
Mach-Zehnder interferometer, the absence of a signal from the mirror F was
attributed to a small deviation of the beam induced by this mirror (as compared
to its width) with the same order of deviation from other mirrors. It was stated
that the nature of the absence of signals from mirrors F and E under these
conditions is the same, which is incorrect, as will be shown below. The accuracy
of absence of signals from mirrors F and E, as well as the dependence of this
accuracy on the feature of the beam, was not analyzed. In the case of
constructive interference in the inner Mach-Zehnder interferometer, a physical
reason for the doubling of the amplitude of modulation of signals from mirrors F
and E as compared to signals from other mirrors was not ascertained. Finally,
the following important question remained unanswered: Are the discussed
paradoxical results the consequences of the features of the used beams (with a
Gaussian profile) and their symmetry or are they valid at any profile of beams?

The aim of this work is to clarify the listed uncertainties using a more direct
general approach differing from that used in \cite{Saldanha2014}. It is shown
that perturbations of the interferometer when performing weak measurements in
the discussed experiment cannot be neglected; moreover, just these perturbations
are entirely responsible for the results of these measurements. The extremely
clear explanation of paradoxical results of the experiment reported in
\cite{Danan2013} will be given within traditional wave theory of light or
quantum mechanics, which implies continuous trajectories of light. Consequently,
an additional concept of discontinuity of trajectories of photons is
unnecessary.

\section{\label{sec:Analisys}ANALYSIS OF THE EXPERIMENT \cite{Danan2013}}

\begin{figure}
	\includegraphics[width=\columnwidth]{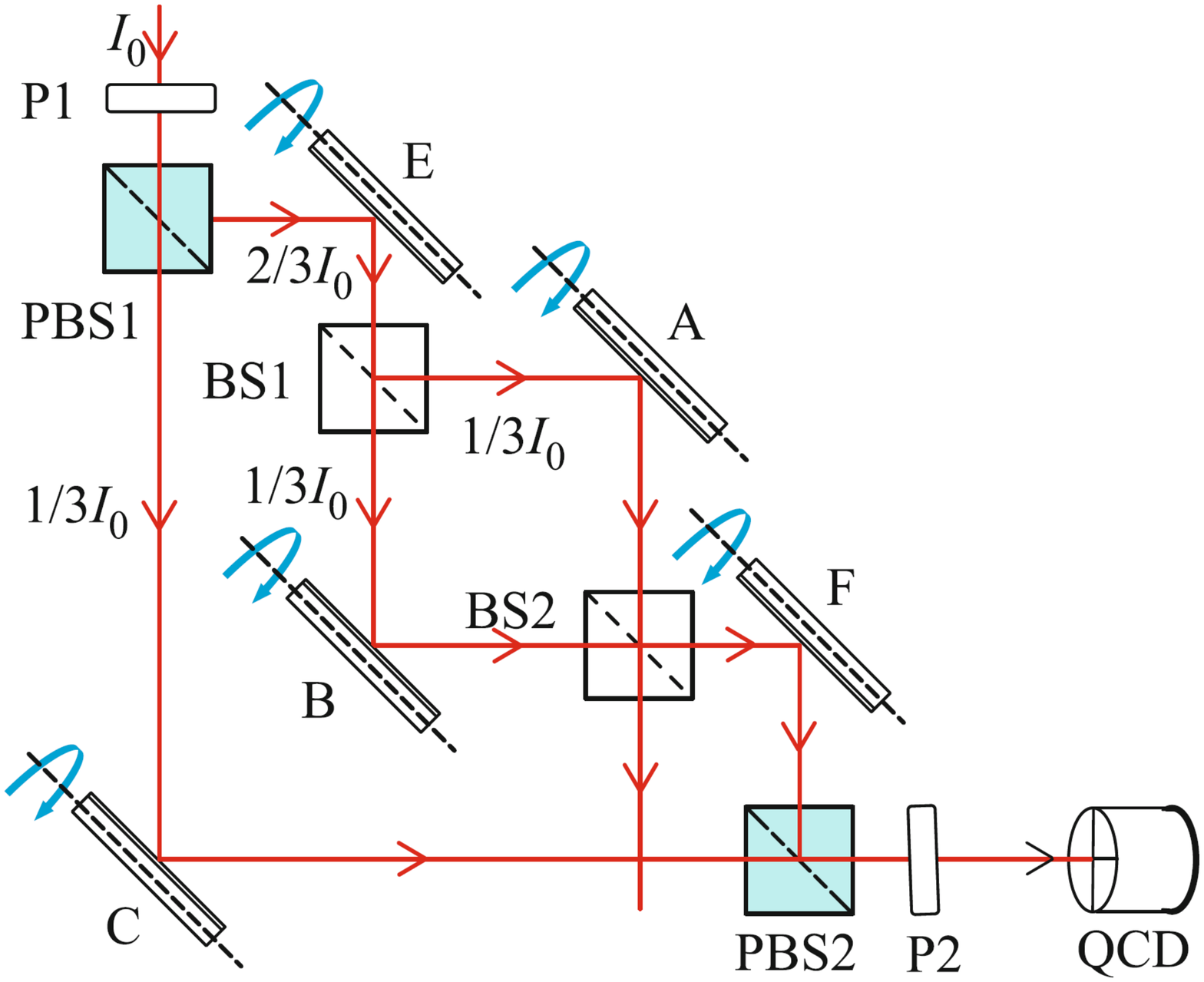}
	\protect\caption{(Color online) Simplified layout of a setup for detection of paths of photons \cite{Danan2013}.   Elements BS1, A, B, and 
		BS2 form the inner Mach-Zehnder interferometer, 
		whereas elements P1, PBS1, C, E, F, PBS2, and P2 form 
		the outer Mach-Zehnder interferometer, where A, B, C, 
		E, and F are mirrors; P1 and P2 are polarizers; BS1 and 
		BS2 are ordinary beam splitters; PBS1 and PBS2 are 
		polarization beam splitters; and QCD is the quad-cell
		photodetector.} 
	\label{fig:setup}
\end{figure}

Figure \ref{fig:setup} shows the simplified layout of the experiment.
Identically tuned polarizers P1 and P2 and polarization beam splitters PBS1 and
PBS2 are placed at the input and output of the nested Mach-Zehnder
interferometer. The polarizer P1 and polarization beam splitter PBS1 are tuned
so that 2/3 and 1/3 of the total intensity of incident light are incident on
mirrors E and C, respectively.
Light beams transmitted through PBS2 and P2 interfere and are detected on the
quad-cell photodetector (QCD). Mirrors A, B, C, E, and F undergo small
vibrations around their horizontal axes with individual frequencies resulting in
vertical shifts of the beam (along the $y$ axis) on the QCD. The frequency power
spectrum of the difference between intensities of light in the upper and lower
parts of the surface of the photodetector is recorded. The ratio of frequencies
of vibration of the mirrors to the frequency of light is $\sim10^{-12}$. The
ratio of the angular deflection of the beam ($\sim300$ nrad) caused by the
modulation of mirrors to the diffraction divergence of the beam is
$\sim10^{-4}$. Consequently, the modulation of mirrors can be neglected when
calculating optical paths. Therefore, the total amplitude of the electric field
$E(x,y)$ at the point ($x,y$) of the detector has the form

\begin{align}\label{eq:amplitude}
E(x,y)&=\dfrac{\sqrt{I_{0}}}{3}\left[f(x,y-\delta_{E}-\delta_{A}-\delta_{F})e^{i\varphi_{A}}+ \right. \notag \\
&+f(x,y-\delta_{E}-\delta_{B}-\delta_{F})e^{i\varphi_{B}}+\\
&+\left. f(x,y-\delta_{C})e^{i\varphi_{C}}\right], \notag
\end{align}
where $I_{0}$ is the intensity of light at the input of the
Mach-Zehnder interferometer; $\varphi_{C}$, $\varphi_{A}$
and $\varphi_{B}$ are
the phase increments of light fields transmitted
through mirrors C, A, and B, respectively, at their
motion from the input of the Mach-Zehnder interferometer
to the detector; $\delta_{C}$,
$\delta_{E}$, $\delta_{A}$, $\delta_{B}$, and $\delta_{F}$ are the
vertical shifts of beams of light because of vibrations of
mirrors C, E, A, B, and F, respectively; and $f(x,y)$ is
the amplitude profile of the incident beam of light
along the horizontal and vertical directions normalized
as $\iint f^{2}(x,y)dxdy=1$.

The QCD records the difference $D$ between the integral intensities of light
from regions $y>0$ and $y<0$:
\begin{equation}
D\equiv\iint_{y>0}\left|E(x,y)\right|^{2}dxdy-\iint_{y<0}\left|E(x,y)\right|^{2}dxdy.\label{eq:detec_diff}
\end{equation}

Since all deviations are small ($\delta\ll1$), $f(x,y-\delta)$ can be
represented in the form of the sum of the first two terms of the Taylor series,
which are the unmodulated initial profile and the term linearly modulated in
$\delta$. In this case, the modulated difference between the integral
intensities of light given by Eq. \eqref{eq:detec_diff} can be represented in
the form

\begin{align}\label{eq:detec_diff_expl}
D&=\dfrac{2}{9}I_{0}\left\{ \delta_{C}+\left[\delta_{A}+\delta_{B}+2\left(\delta_{E}+\delta_{F}\right)\right]\left[1+\cos\left(\phi_{AB}\right)\right]+\right. \notag \\ &+\left[\delta_{A}+\delta_{C}+\delta_{E}+\delta_{F}\right]\cos\left(\phi_{AC}\right)+\\ &+\left.\left[\delta_{B}+\delta_{C}+\delta_{E}+\delta_{F}\right]\cos\left(\phi_{BC}\right) \right\} \int{f^{2}(x,0)dx},  \notag
\end{align}
\[ \phi_{ij}\equiv \varphi_{i}-\varphi_{j}. \]
When the profile of the beam of light is symmetric, i.e., $f(x,-y)=f(x,y)$,
which corresponds to the conditions of the experiment reported in
\cite{Danan2013}, Eq. \eqref{eq:detec_diff_expl} is the total signal on the QCD
photodetector. In the case of an arbitrary profile, there is also an additional
constant component. The first line in the curly brackets in Eq.
\eqref{eq:detec_diff_expl} is the sum of self-interferences of the modulated and
unmodulated parts of all beams of light moving through different arms of the
composite Mach- Zehnder interferometer. Here, the coefficient 2 of the term
$\left(\delta_{E}+\delta_{F}\right)$ appears because mirrors E and F modulate
beams of light moving through both arms of the inner Mach-Zehnder
interferometer. The second and third lines in the curly brackets in Eq.
\eqref{eq:detec_diff_expl} are due to the interference of beam C with beams A
and B, respectively.

In the case of constructive coherence of beams of light passing through mirrors
A, B, and C, i.e.,
$\varphi_{A}=\varphi_{B}=\varphi_{C}$, the expression in the curly brackets in
Eq. \eqref{eq:detec_diff_expl} becomes the form
\begin{equation}
\left\{ \cdots\right\} =3\left\{ \delta_{A}+\delta_{B}+\delta_{C}+2\left(\delta_{E}+\delta_{F}\right)\right\} .\label{eq:constr_interf}
\end{equation}

The power spectrum of the difference signal given by Eqs. \eqref{eq:detec_diff}
and \eqref{eq:detec_diff_expl} rather than the signal itself was detected in the
experiment reported in \cite{Danan2013}. For this reason, according to Eq.
\eqref{eq:constr_interf}, the intensity of spectral components of the signal
given by Eq. \eqref{eq:detec_diff_expl} that correspond to vibrations of mirrors
E and F is four times larger than the others (see Fig. 2a in \cite{Danan2013}).

When the phase $\varphi_{B}$ is changed by $\pi$ in the inner Mach-Zehnder
interferometer ($\varphi_{A}=\varphi_{C}=\varphi_{B}\pm\pi$), which corresponds
to the complete destructive interference of light at its output, the expression
in the curly brackets in Eq. \eqref{eq:detec_diff_expl} becomes the form
$\left\{ \cdots\right\} =\left\{ \delta_{A}-\delta_{B}+\delta_{C}\right\}$. In
this case, the intensities of spectral components of the signal in the detector
that are due to vibrations of mirrors A, B, and C will be identical (see Fig. 2b
in \cite{Danan2013}). At the same time, the spectral components at the
frequencies of vibrations of mirrors E and F are absent. The authors of
\cite{Danan2013} interpreted this extraordinary result as the discontinuity of
trajectories of photons, which are as if present in the inner Mach-Zehnder
interferometer but are absent at its input and output. The authors believe that
a simple and intuitively clear explanation of such an extraordinary result is
possible only within the two-state vector formalism of quantum mechanics
\cite{Aharonov1964,Aharonov1990}. Within the framework of this formalism, it is
asserted that photons are present only where there is both a direct quantum wave
(from the source) and an inverse one (from the detector) exist (see Fig. 3 in
\cite{Danan2013}).

If light is blocked immediately behind mirror F, the detector naturally records
a signal only from light in the lower arm of the outer Mach-Zehnder
interferometer. In this case, only the last term remains in the square brackets
in Eq. \eqref{eq:amplitude} and only the first term in the curly brackets in Eq.
\eqref{eq:detec_diff_expl} is nonzero: $\left\{ \cdots\right\} =\left\{
\delta_{C}\right\} $.
\begin{figure}[t]
	\includegraphics[width=\columnwidth]{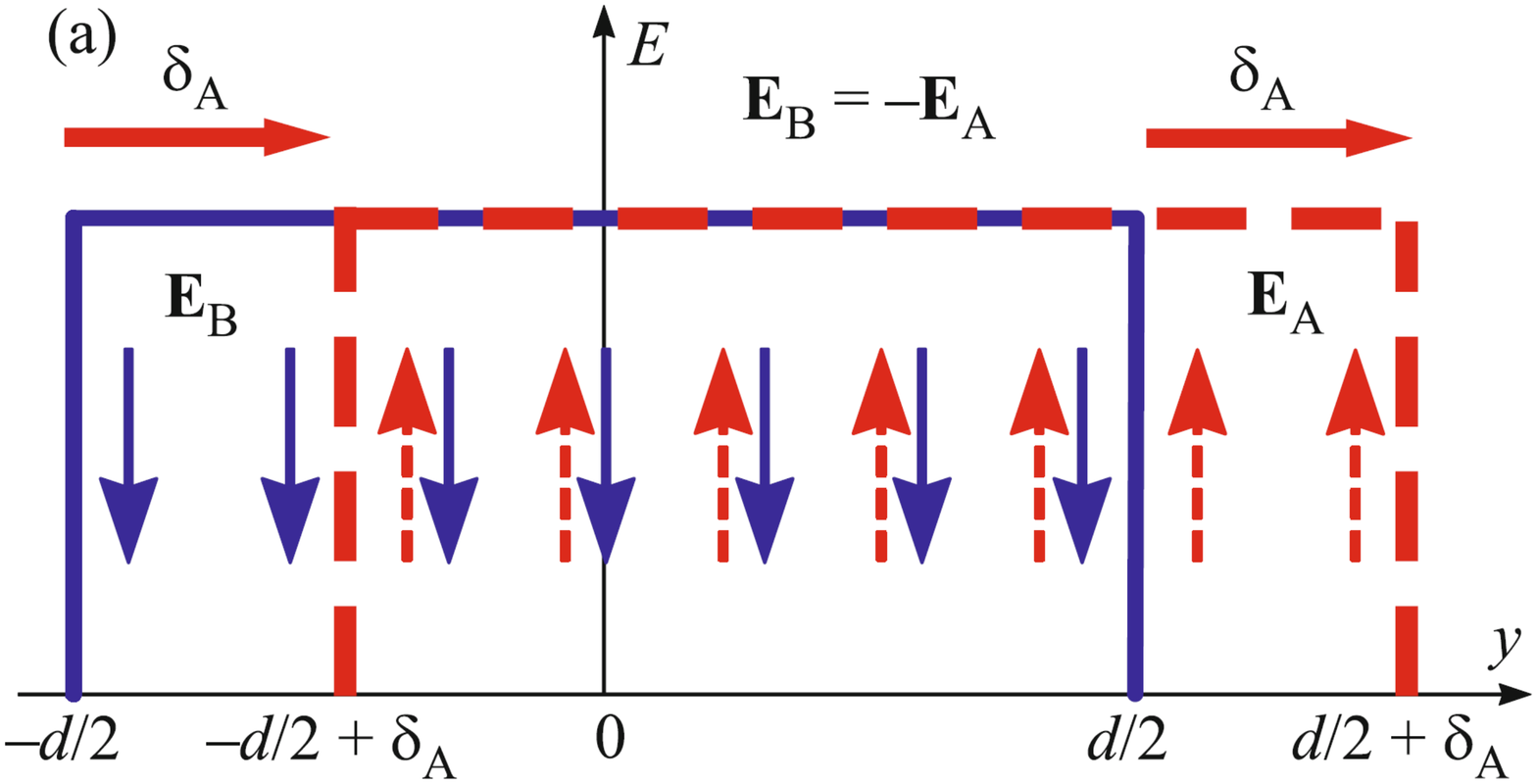}\label{fig: profile_on_mirror_F_a}
	\includegraphics[width=\columnwidth]{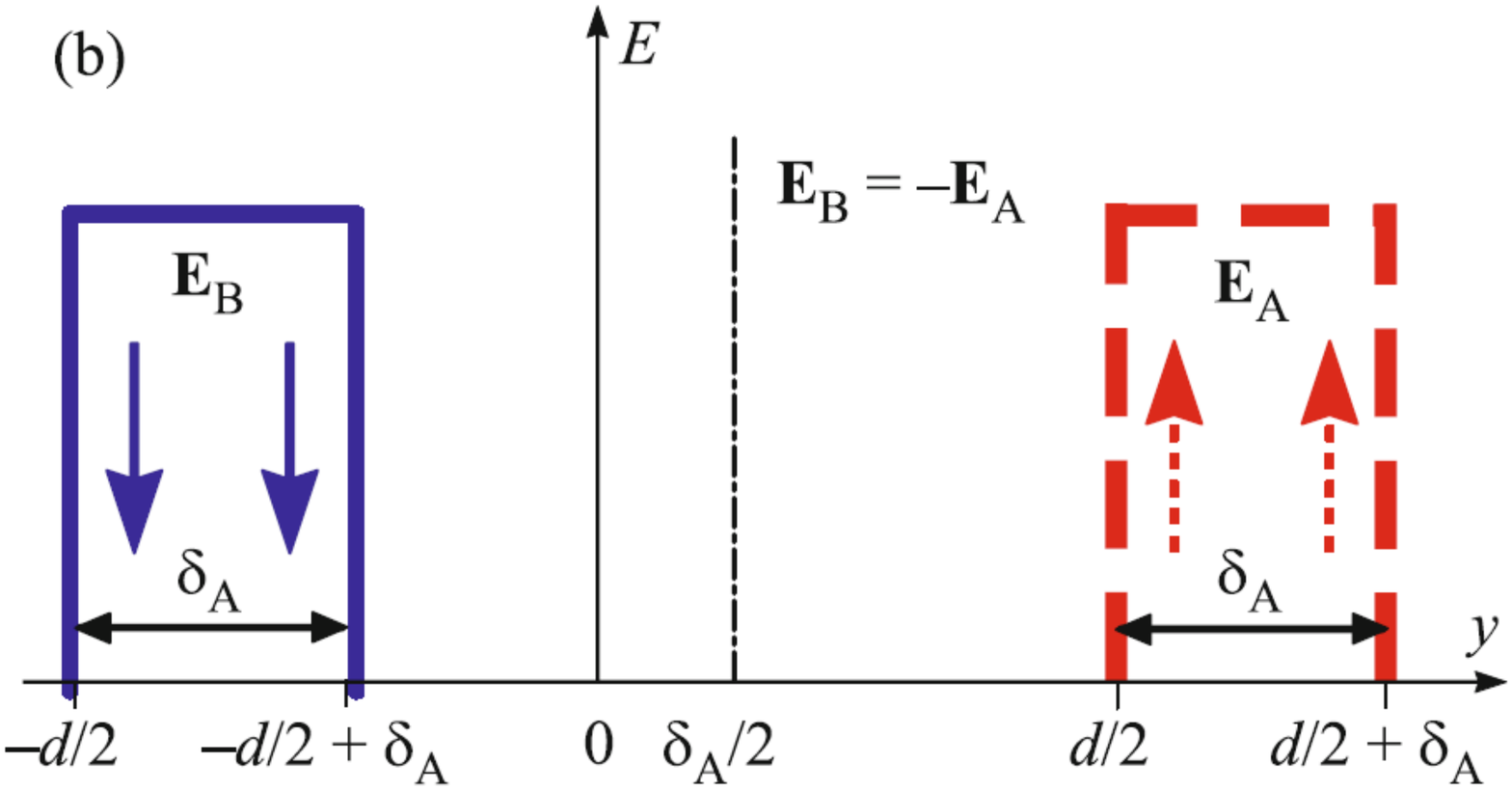}\label{fig: profile_on_mirror_F_b}
	\protect\caption{(Color online) Transverse distribution 
	        of the amplitude of light at the output of the inner Mach-Zehnder
		interferometer (in front of mirror F, see Fig. \ref{fig:setup}), tuned to
		complete destructive interference of rectangular beams
		with the width $\mathrm{d}$ passing through it. (a) Mutual transverse
		arrangement of the beam passing through mirrors B and A
		(the case of the deviation of only one beam by  $\delta_{A}$
		caused
		by deflection of mirror A is shown for simplicity). (b)
		Result of their destructive interference. The vertical arrows
		mark the instantaneous positions of electric-field vectors
		of the corresponding beams.}
	\label{fig: profile_on_mirror_F}
\end{figure}

If light is blocked between mirror C and polarization splitter PBS2, the last
term in the square brackets in Eq. \eqref{eq:amplitude} is absent and only the
second term in the curly brackets in Eq. \eqref{eq:detec_diff_expl} is nonzero:
$\left\{ \cdots\right\} =\left\{
\left[\delta_{A}+\delta_{B}+2\left(\delta_{E}+\delta_{F}\right)\right]\left[1+\cos\left(\phi_{AB}\right)\right]\right\}
$. In the case of the complete destructive interference of light at the output
of the inner Mach-Zehnder interferometer
($\varphi_{A}=\varphi_{C}=\varphi_{B}\pm\pi$), the expression in the curly
brackets vanishes; i.e., the intensities of spectral components of the signal of
the detector that are caused by vibrations of all mirrors are zero (see Fig. 2ñ
in \cite{Danan2013}).

Such paradoxical results in the complete destructive interference of light
($\varphi_{A}=\varphi_{C}=\varphi_{B}\pm\pi$) at the output of the inner
Mach-Zehnder interferometer can be simply and clearly explained within both the
wave theory of light and traditional quantum mechanics.

Indeed, spectral components with the frequency of vibration of mirror E are
absent because the vibration of this mirror identically displaces beams of light
passing through mirrors A and B. For this reason, interference of these beams at
the output of the inner Mach-Zehnder interferometer does not change at the
displacement of mirror E. Consequently, light at the output of this Mach-Zehnder
interferometer is absent both in the case of immobile mirror E and in the case
of its vibrations.
\begin{figure}
	\includegraphics[width=\columnwidth]{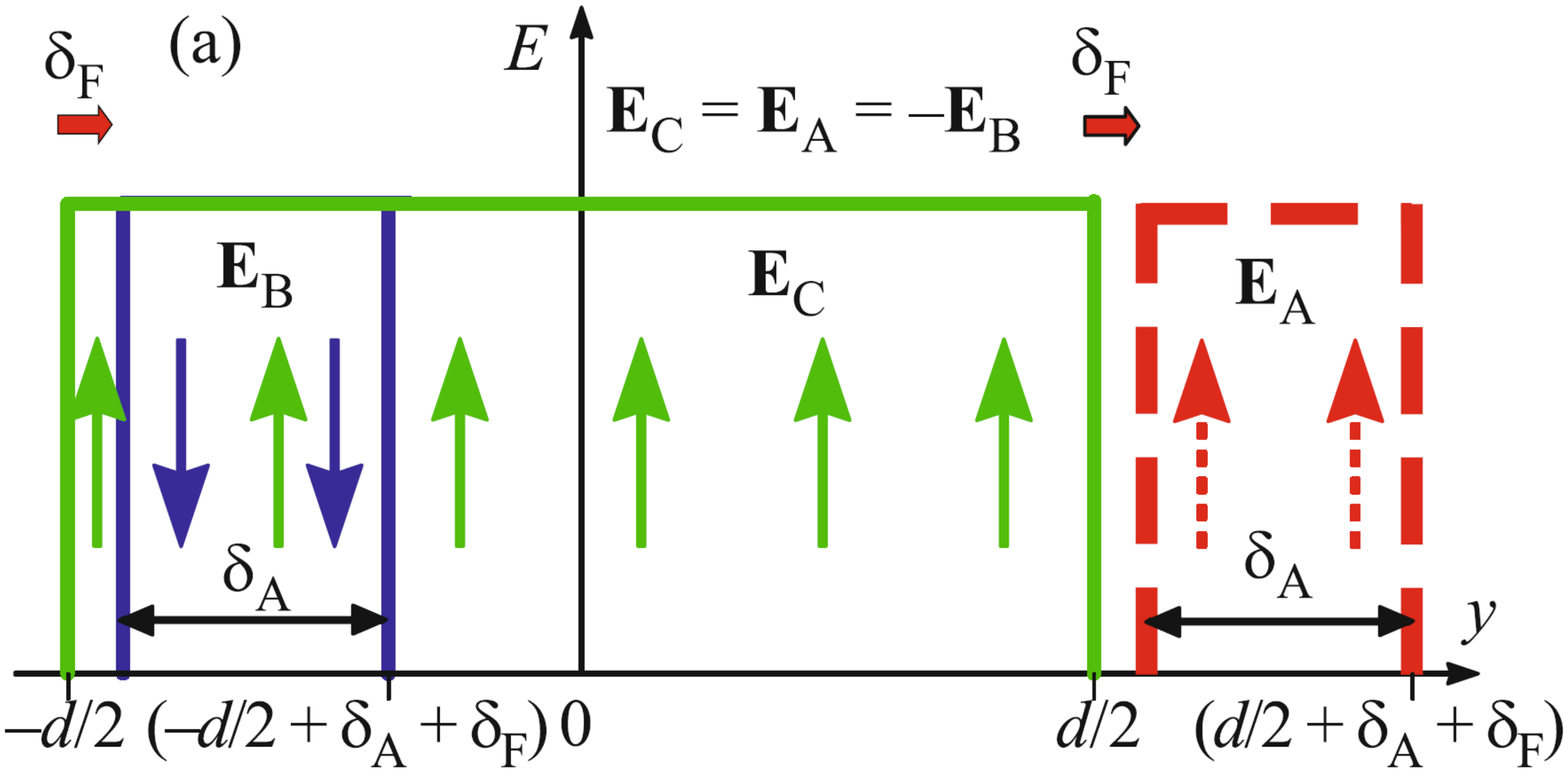}\label{fig: profile_on_detector_a}
	\includegraphics[width=\columnwidth]{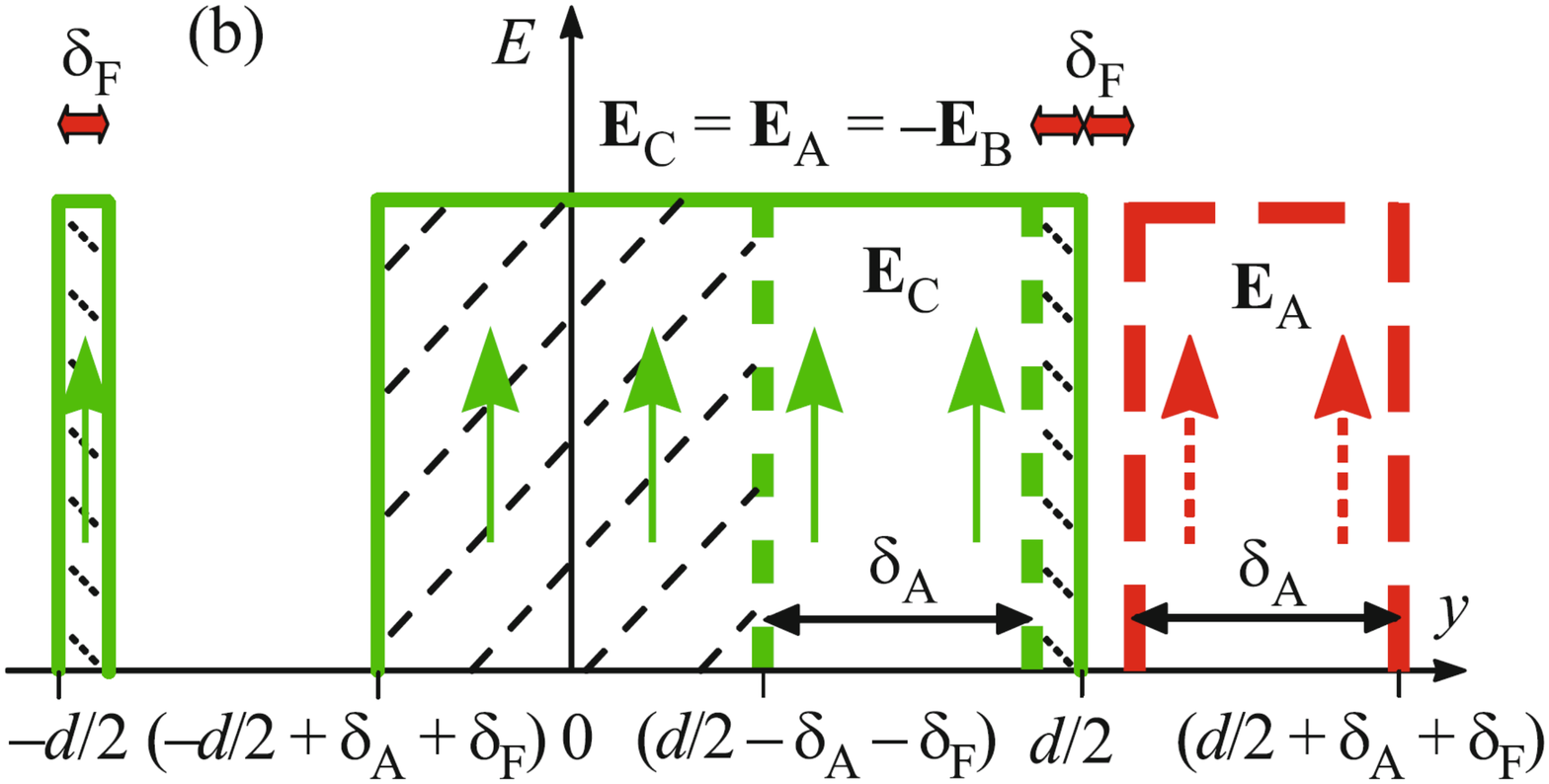}\label{fig: profile_on_detector_b}\\
	\protect\caption{(Color online) Transverse distribution
	        of the amplitude of light in front of the photodetector. For simplicity,
		the direction of the polarization of fields in this figure and
		in Fig. \ref{fig: profile_on_detector_final} remains unchanged (in reality, polarizer P2
		changes this direction). (a) Mutual transverse arrangement
		of the beams transmitted through mirrors C and F on the
		photodetector. The latter mirror additionally displaces a
		beam by $\delta_{F}$. (b) Result of their destructive interference.
		The shaded symmetric regions do not contribute to a signal of the quad-cell photodetector QCD.}
	\label{fig: profile_on_detector}
\end{figure}

On the contrary, at the displacement of any of mirrors A and B, destructive
interference of beams of light from these mirrors at the output of the inner
Mach-Zehnder interferometer is incomplete because the amplitudes of interfering
beams are no longer equal to each other at any point of the cross section of the
outgoing beam. More precisely, in the case of $\delta\ll 1$ and a symmetric
profile of beams, this destructive difference is an antisymmetric function of
$y$ in the first order in $\delta$. In turn, a change in this antisymmetric
function caused by small vibrations of mirror F has the second order of
smallness and is a symmetric function of $y$. A fraction of light incident on
the photodetector, which is modulated with the frequency of deviations of mirror
F, is the result of interference of the modulated fraction of light from mirror
F and the unmodulated fraction of light from mirror C. Both of these components
are symmetric functions of $y$. Consequently, their interferential sum is also
symmetric with an accuracy up to the third order in $\delta$. This is the reason
why the QCD does not record it. In the case of an arbitrary profile of the beam,
the signal from mirror F is not detected with an accuracy of the second order of
smallness.
\begin{figure}
	\includegraphics[width=\columnwidth]{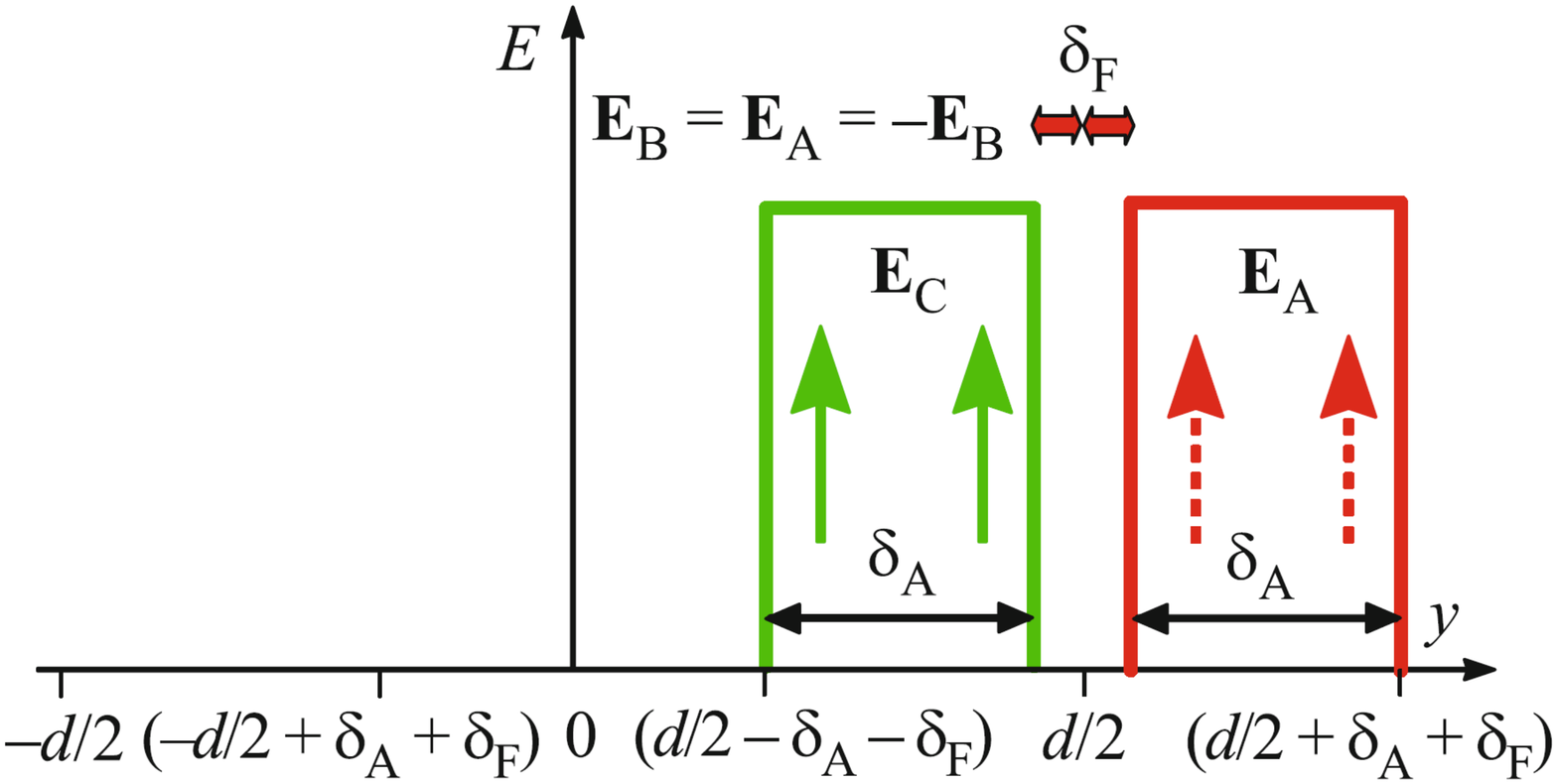}
	\protect\caption{(Color online) Detected part of the distribution of amplitude of light on the QCD detector. The amplitude obviously depends only on the deflection of mirror A. The 	deflection of mirror F does not affect a detected signal.}
	\label{fig: profile_on_detector_final}
\end{figure}
The signal detected in this case is due to both the modulated part of the light
from mirror C and the result of interference of its unmodulated part fraction
with the light coming from the inner Mach-Zehnder interferometer and being the
differential of the profile of the initial light beam. The result of this
interference is equivalent (with an accuracy up to the second order regard to
$\delta$) to the initial unmodulated beam of light from mirror C displaced at
the QCD on the shift ${\delta}_A$ $-{\delta}_B$ of the beams of light
transmitted through the inner Mach-Zehnder interferometer.

A similar physically transparent interpretation can be given for the absence of
signals from all mirrors at $\varphi_{A}=\varphi_{B}\pm\pi$ and at blocking of
light between mirror C and polarization splitter PBS2. The absence of a signal
from mirror C is obvious and the absence of signals from mirrors E and F was
discussed above. Signals from mirrors A and B are absent because the amplitude
profile of light in this case is an antisymmetric function of $y$ (with an
accuracy of the second order in $\delta$) not only at the output of the inner
Mach-Zehnder interferometer but also on the QCD. At the same time, the QCD
measures the difference of integral intensities given by Eq.
\eqref{eq:detec_diff}, which vanishes in this case (with an accuracy of the
third order in $\delta$). In the case of an arbitrary profile of the beam, any
signal from mirrors A and B is not detected with an accuracy of the second order
of smallness.

An important feature of discussed signals should be emphasized. Such weak
measurements of the presence of photons in various places of the embedded
Mach-Zehnder interferometer are completely due to the interference between the
modulated and unmodulated parts of beams of light, because modulated parts of
the amplitude given by Eq. \eqref{eq:amplitude}, as well as detected signals
given by Eq. \eqref{eq:detec_diff_expl}, are proportional to small deviations
$\delta$. In the general case, each modulated part interferes with all
unmodulated parts.

The marked ``interference'' feature is an inherent part of any weak measurements
\cite{Duck1989}. To illustrate the indicated interference features, it is
convenient and instructive to consider beams of light with a symmetric step
profile where the amplitude of light is constant at any point of the cross
section of the beam. A remarkable feature of such a profile is that all above
results are exact for it if the total deviation of the beam is no more than its
half-width. The reason is that the difference between the displaced and initial
beams with such a profile is an exactly antisymmetric function on ($y-\delta/2$)
at an arbitrary transverse deviation less than the half-width of the beam.
Figures \ref{fig: profile_on_mirror_F} -- \ref{fig: profile_on_detector_final}
show the amplitude profiles of beams of light in various places of the embedded
Mach-Zehnder interferometer and on the QCD in the case of complete destructive
interference of light at the output of the inner Mach-Zehnder interferometer
(for clarity, it is accepted that mirror C is at rest). It is clearly seen (in
Figs. \ref{fig: profile_on_detector} and \ref{fig: profile_on_detector_final})
that a nonzero integral difference signal specified by Eq. \eqref{eq:detec_diff}
appears only at the interference of light from the inner Mach-Zehnder
interferometer and unmodulated light from mirror C. A signal from mirror F does
not appear for the same reason that was discussed above for an arbitrary
symmetric profile of the beam and small deviations $\delta$. Indeed, the
difference between the amplitudes at the output of the inner Mach-Zehnder
interferometer and behind mirror F is a symmetric function of
($y-\delta_{A}/2-\delta_{F}/2$). The detector records the result of the
interference of this difference and the symmetric unmodulated component of light
from mirror C. The integral contribution of this interference vanishes if the
total deviation $\delta_{A}+\delta_{F}$ is no more than the half-width of the
beam.

\section{\label{sec:Conclusions}CONCLUSIONS}\
To summarize, it has been shown that paradoxical results of the experiment
reported in \cite{Danan2013}, which were interpreted by the authors as the
\emph{discontinuity} of possible trajectories of photons, can be simply and
clearly explained within the traditional concepts of the wave and quantum
natures of light, \emph{which are based on the continuity of all possible paths
of photons}. It has been established that extraordinary signals detected in
\cite{Danan2013} are completely due to the perturbation of destructive
interference of light in the inner Mach-Zehnder interferometer.

It has been shown that the absence of signals from mirrors at the input and
output of the inner Mach-Zehnder interferometer at its destructive tuning is due
not to \emph{discontinuity of trajectories of light (photons)}, as was stated in
[8], but to \emph{the used method of detection of paths of photons} (harmonic
deviation of the mirrors of the interferometer for a small deviation of
trajectories of light). The modification of the scheme can reveal "traces" of
photons at the indicated places. In particular, the path of photons through
mirror E can be detected if this mirror does not deflect light but modulates its
polarization and a birefringent plate and a polarizer are placed in one of the
arms of the inner Mach-Zehnder interferometer.

Finally, the performed analysis has indicated that it is unnecessary to
introduce a new physical concept of \emph{discontinuity} of possible
trajectories of photons, which was proposed by the authors in the discussed work
on the basis of the results of the performed experiment and their interpretation
within their treatment of the two-state vector formalism.

I am grateful to V.A. Sorokin for stimulating discussions. This work was
supported by the Government of the Russian Federation (project no. 01201372518,
Program of Basic Research for the State Academies of Sciences).

\end{document}